\documentclass[man, 12pt, floatsintext]{apa7}

\usepackage[american]{babel}
\usepackage{graphicx}
\usepackage{subcaption}
\usepackage{comment}

\usepackage{csquotes} 
\usepackage[style=apa,sortcites=true,sorting=nyt,backend=biber]{biblatex}
\DeclareLanguageMapping{american}{american-apa} 
\usepackage[T1]{fontenc} 
\usepackage{mathptmx} 

\title{\Large{Impact of whole-body vibrations on electrovibration perception varies with target stimulus duration}}

\shorttitle{Electrovibration in vehicle cockpits} 
\authorsnames[1, 2, 1]{Jan D.~A. Vuik, Daan M. Pool,  Yasemin Vardar}
\authorsaffiliations{{Department of Cognitive Robotics, Faculty of Mechanical Engineering, Delft University of Technology, Delft, The Netherlands},{Department of Control and Operations, Faculty of Aerospace Engineering, Delft University of Technology, Delft, The Netherlands}}

\authornote{\addORCIDlink{Daan M. Pool}{0000-0001-9535-2639}\\
\addORCIDlink{Yasemin Vardar}{0000-0003-2156-1504}\\

Manuscript Type: Research Article\\ 
Word Count: 4450 \\ 
Corresponding Author: Yasemin Vardar, Department of Cognitive Robotics, Delft University of Technology, Mekelweg 2, 2628 CD, Delft, The Netherlands\\ 
Funding Source: This publication is part of the project ``Right Touch Right Time: Future In-vehicle Touchscreens (FITS)'' with project number 20624 of the Open Technology Research Program financed by the Dutch Research Council (NWO). YV is also funded by the project (number 19153) of the NWO research program VENI.\\
Acknowledgments: The authors thank Ferdinand Postema for his help in designing the experimental setup.}

\abstract{\textbf{Objective:} This study explores the impact of whole-body vibrations induced by external vehicle perturbations, such as aircraft turbulence, on the perception of electrovibration displayed on touchscreens.\\
\noindent\textbf{Background:} Electrovibration holds promise as a technology for providing tactile feedback on future touchscreens, addressing usability challenges in vehicle cockpits. However, its performance under dynamic conditions, such as during whole-body vibrations induced by turbulence, still needs to be explored.\\
\noindent\textbf{Method:} We measured the absolute detection thresholds of 15 human participants for short- and long-duration electrovibration stimuli displayed on a touchscreen, both in the absence and presence of two types of turbulence motion generated by a motion simulator. Concurrently, we measured participants' applied contact force and finger scan speeds.\\
\noindent\textbf{Results:} Significantly higher (38\%) absolute detection thresholds were observed for short electrovibration stimuli than for long stimuli. Finger scan speeds in the direction of turbulence, applied forces, and force fluctuation rates increased during whole-body vibrations due to biodynamic feedthrough. As a result, turbulence also significantly increased the perception thresholds, but only for short-duration electrovibration stimuli.\\
\noindent\textbf{Conclusion:} The results reveal that whole-body vibrations can impede the perception of short-duration electrovibration stimuli, due to involuntary finger movements and increased normal force fluctuations.\\
\noindent\textbf{Application:} Our findings offer valuable insights for the future design of touchscreens with tactile feedback in vehicle cockpits.}

\keywords{Tactile displays, touchscreens, haptics, psychophysics, biodynamics\\
\begin{center}
\textbf{Prec\'is}\\
\end{center}
This study investigates the impact of external vehicle perturbations, like aircraft turbulence, on electrovibration perception on touchscreens. Whole-body vibrations were found to cause perturbed finger movements and normal force fluctuations, leading to elevated thresholds for short-duration stimuli. These findings should guide the design of tactile feedback touchscreens for vehicle cockpits.
}

\begin{document}    

\maketitle 

\section{Introduction}
\label{sec:introduction}

In today's digital era, touchscreens have become indispensable and are integrated into various electronic devices like smartphones, tablets, laptops, kiosks, and digital information panels. Their presence has also surged in vehicle cockpits across automobiles, aircraft, and vessels as they offer a multitude of advantages over traditional buttons and knobs~\parencite{Ahmad_Langdon_Godsill_2018, VanZon2020}. Touchscreens can effortlessly incorporate vast amounts of information, updated through reconfigurable graphical user interfaces (GUIs) without needing to rewire mechanical controls. Additionally, they foster intuitive interactions through pointing gestures, elucidating their prevalence in contemporary vehicle cockpits.

However, challenges arise with the widespread use of touchscreens in these environments. One notable issue is the absence of tactile or aural feedback. Unlike traditional knobs and buttons that offer feedback through force or sound~\parencite{Wang_Wang_Chen_2018}, touchscreens require users to confirm actions through alternative means, such as visually checking if a button has been pressed. This lack of tangible feedback can pose safety risks, particularly when users divert their attention to the touchscreen instead of the road, potentially leading to hazardous situations in traffic.

Moreover, using touchscreens during external perturbations, such as turbulence in aircraft, bumpy roads in cars, or waves in ships, presents difficulties. These perturbations and vibrations induce relative motions between the user's finger and the screen due to \emph{biodynamic feedthrough} \parencite{venrooij2013bdft}, hindering touchscreen use~\parencite{cockburn:hal-02649024, dodd2014touch, Coutts2019}. Various studies indicate that environmental vibrations negatively impact user performance across different input devices \parencite{mcleod1980influence, hill2005soldier, NARAYANAMOORTHY20112263}, and that this effect extends to touchscreens in vehicle cockpits \parencite{cockburn:hal-02649024, dodd2014touch, Coutts2019, Liu_ergonomic_evaluation, Khoshnewiszadeh2021680}. Various methods have been explored to address these challenges, including larger buttons, increased spacing, visual and auditory feedback, and additional physical features on the screen. However, none provide a complete solution to this problem~\parencite{dodd2014touch, Coutts2019, Liu_ergonomic_evaluation, visual_and_auditory_feedback, Lancaster_Mers_Rogers_Smart_Whitlow_2011}. 

Surface haptics, specifically electrovibration, emerges as a potential solution to mitigate the decrease in task performance during external perturbations. Electrovibration modulates perceived friction through induced electrostatic forces between a finger and a high-voltage supplied capacitive touchscreen~\parencite{teslatouch}. It has shown positive effects on touchscreen user performance, enhancing accuracy, efficiency, and task completion times during pan gestures and dragging tasks~\parencite{liupangestures, zhangharrisondragging}. Exploring its application in button or ridge rendering could improve task performance, as demonstrated in earlier work conducted with other types of surface haptic displays~\parencite{corentin2022eyesoff}.

Despite its potential advantages for improving touchscreen interactions in vehicle cockpits, little is known about how electrovibration technology can be utilized during external vehicle perturbations (e.g., sudden impacts, bumpy roads, waves, and turbulence). So far, this technology has been developed and tested exclusively in conditions where users interact with devices on a static table. Nonetheless, it is well-known that physical perturbations directly interfere with perceived tactile sensations due to undesirable masking effects, such as a reduced perceived intensity or not noticing the tactile stimulus~\parencite{verrillo1985vibrotactile, ryu_mechanical_vibration, vardar2018masking, Jamalzadehremotemasking}. Moreover, changes in finger contact force and speed affect the generated tactile stimulus due to induced variations in the finger contact area~\parencite{ayyildiz2018contact}, stick-slip behavior~\parencite{ozdamar2020step}, and the air gap between the touchscreen and the skin~\parencite{vardar2021motion, shultz2018gap}. Such changes to the finger-touchscreen interface significantly affect generated tactile stimuli (e.g., forces) and what users feel~\parencite{vardar2021motion}. Understanding how these factors affect electrovibration forces and perception could be leveraged to counteract them, for example by carefully adapting tactile stimuli based on the (measured) signal characteristics of the perturbations.

Here, we investigate the effect of whole-body vibrations in the vertical direction due to external vehicle perturbations (e.g., turbulence) on the perception of electrovibration displayed on touchscreens. For this goal, we conducted psychophysical experiments where we tested the detection thresholds of human participants for short- and long-duration electrovibration stimuli in the absence and presence of whole-body vibrations.  Concurrently, we measured the applied contact force and speed, recognizing their influence on finger contact area and electrovibration perception.

\subsection{Hypotheses}\label{subsection:hypotheses}
We expect a shorter-duration electrovibration stimulus to be more challenging to feel, resulting in higher perceptual thresholds. Next, we anticipate that the whole-body vibrations (e.g., external vehicle perturbations) negatively affect the perception of electrovibration, resulting in higher thresholds. Moreover, we expect that the vertical finger scan speed, the applied normal force, and the rate of change in this force will increase as a result of involuntary limb movement due to vertical whole-body vibrations. 

\section{Methods}
To test our hypothesis, we measured the absolute detection thresholds of human participants for electrovibration stimuli, both in the presence and absence of whole-body vibrations (e.g., turbulence). The experimental procedure was conducted following the Declaration of Helsinki and approved by TU Delft's Human Research Ethics Committee with case number 3280. 

\subsection{Participants}
The psychophysical experiments were conducted with fourteen male and four female participants with an average age of 25.2 years with a standard deviation of 4.1 years. All participants were right-handed. They read and signed an informed consent form before participating in the experiment. 

\subsection{Experimental setup}
\begin{figure}[h!]
\captionsetup[subfigure]{justification=centering}
    \centering
    \begin{subfigure}{0.49\linewidth}
        \includegraphics[width=1\linewidth]{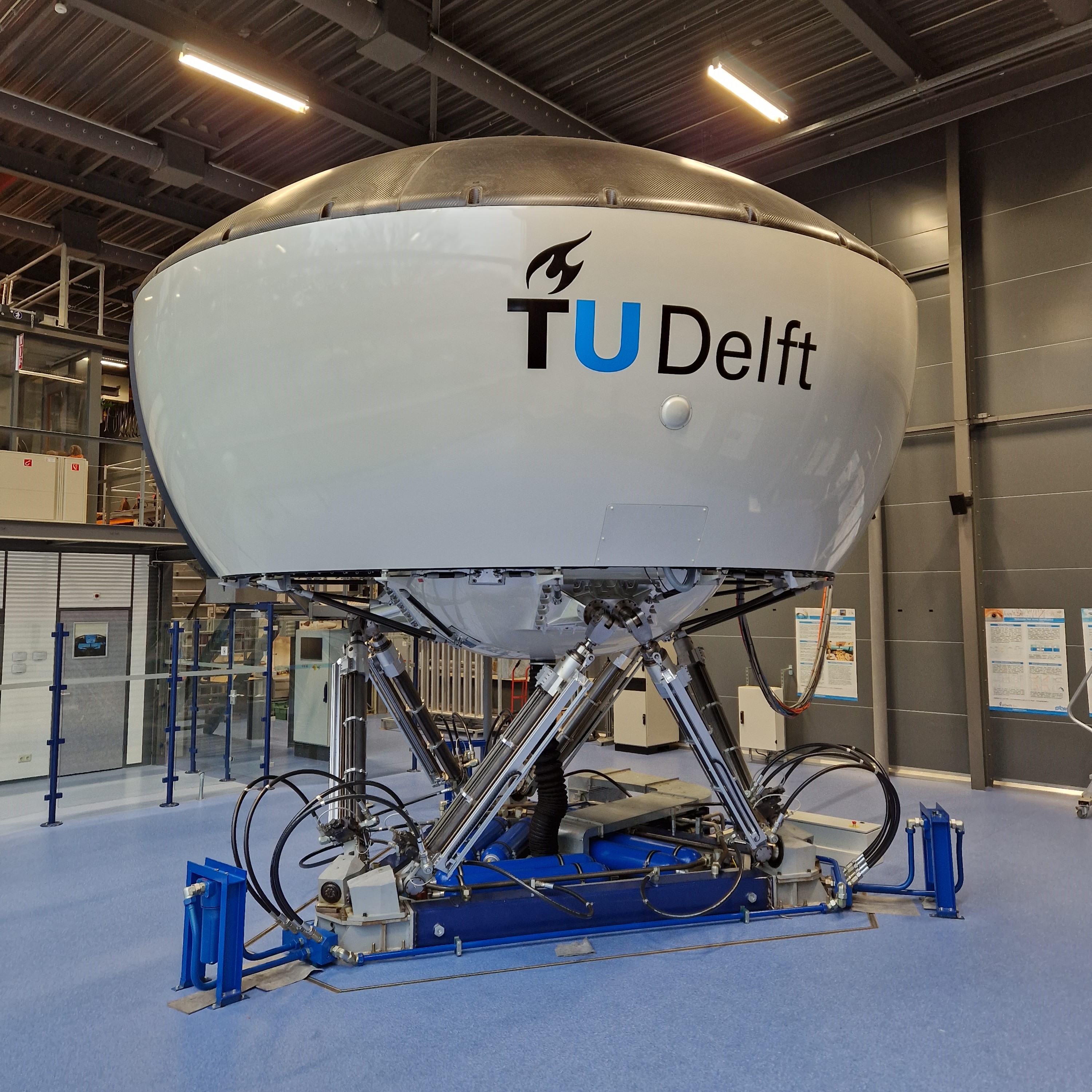}
        \caption{}
        \label{fig:simona}
    \end{subfigure}
    \begin{subfigure}{0.49\linewidth}
        \includegraphics[width=1\linewidth]{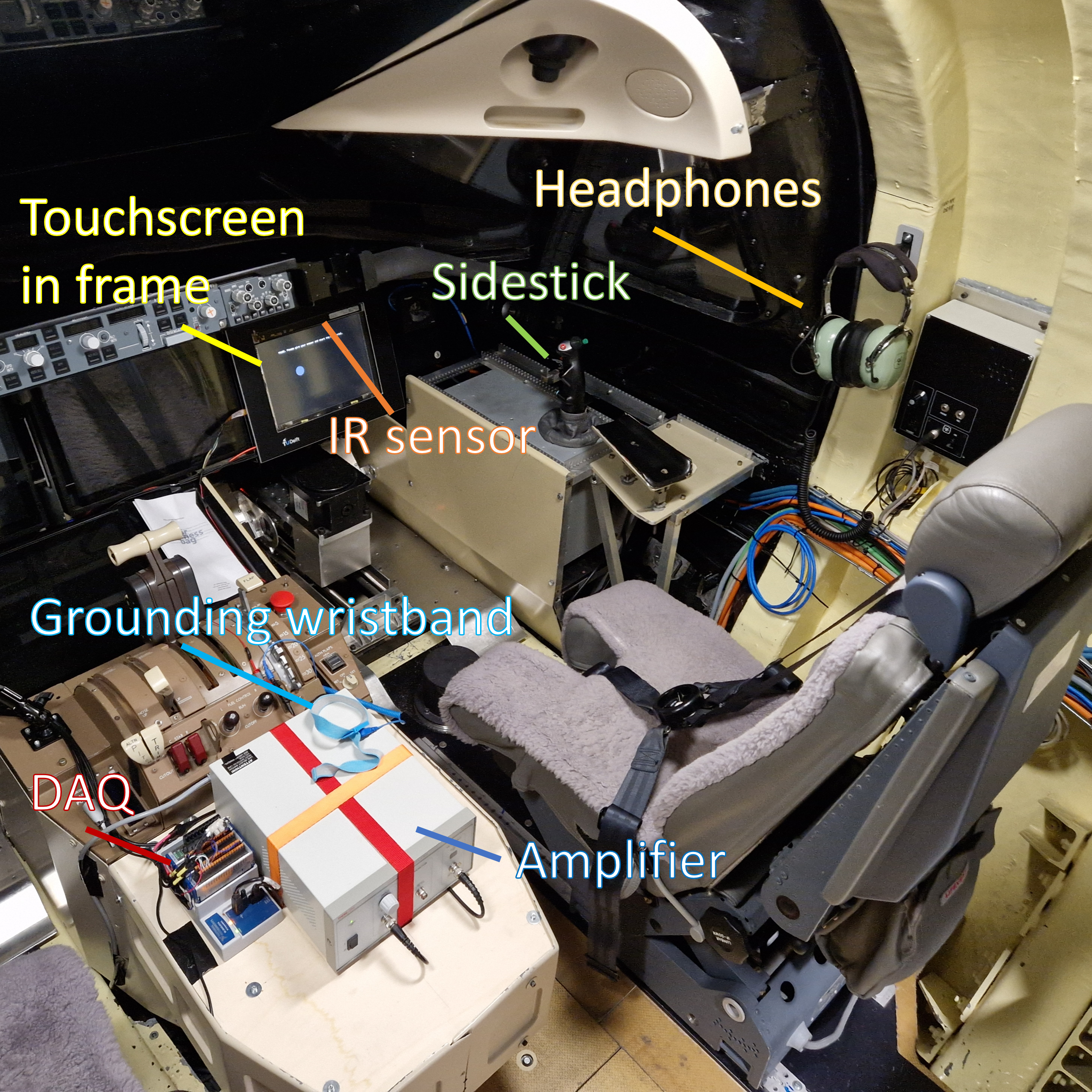}
        \caption{}
        \label{fig:setup}
    \end{subfigure}
    \caption{The experimental setup. (a) The outside view of SIMONA research simulator, (b) the simulator's cabin with experimental equipment.}
\end{figure}

The experiments were conducted in the SIMONA Research Simulator at the Faculty of Aerospace Engineering of TU Delft, see Figure~\ref{fig:simona}. The simulator's 6 degrees-of-freedom hexapod motion system \parencite{Berkouwer2005MeasuringSystem} was used to generate vertical whole-body vibrations simulating aircraft turbulence. 

During the experiments, the participant sat in the right seat of the simulator and was strapped in by the five-point harness, see Figure~\ref{fig:setup}. In front of the participant, a touchscreen (SCT3250, 3M Inc.) was mounted in a custom-made frame that covered the regular primary flight display (ProLite TF1534MC-B1X, Iiyama Corp.). The touchscreen had an 18-degree angle to the vertical plane in this configuration. The original flight display under the frame showed a graphical user interface to guide the participants during the experiments. Within the frame, the touchscreen was attached to four force sensors (FSG020WNPB, Honeywell Inc.) placed at the corners via double-sided tape. These sensors measured the normal force exerted on the touchscreen by the user's finger. A data acquisition board (NI-9205, NI Inc.) collected force sensor data at a sampling rate of 2~kHz. An infrared position sensor (NNAMC2300PCEV, Neonode Inc.) mounted on the top edge of the touchscreen measured the participant's finger position and velocity. The frame encapsulated the edges of the touchscreen, the sensors, and the wiring to keep the setup in place while the simulator was moving and as a safety measure for the participants. During the experiments, the participants could move the seat to the front to be comfortable while interacting with the touchscreen.   

The electrovibration stimuli were generated by applying voltage signals to the touchscreen. These signals were generated through another data acquisition card (NI-9264, NI Inc.) and then augmented by a high-voltage amplifier (HVA200, Thorlabs Inc.). The infrared position sensor and the data acquisition card were connected via USB cable extenders to a computer outside the simulator, in the simulator's control room. The participants wore an anti-static wristband on their non-dominant wrist for grounding and a noise-cancelling aviation headset (H10-66XL, David Clark Inc.) with which they could communicate and hear instructions from the experimenter in the control room. During the experiment, the participants listened to aircraft engine noise to mask any auditory cues from the simulators' motion system. The participants entered their responses through the side stick, see Figure~\ref{fig:setup}; they used the red and green side buttons to select a stimulus interval and the trigger button in the front to initiate the start of a trial. This sidestick was on the right side of the seat, which meant all participants had to use the same hand to press the buttons and touch the screen. A delay of one second was implemented between pressing the button and starting the trial to avoid losing touch data. 

\subsection{Stimuli}
\subsubsection{Target stimulus (electrovibration)}
The target stimulus was an oscillating electrostatic force (i.e., electrovibration) generated by applying a sinusoidal voltage signal at 100~Hz without a DC offset to the conductive layer of the touchscreen. As demonstrated in previous studies on electrovibration perception~\parencite{vardar2017waveform, vardar2018masking, vardar2021motion}, this input voltage generates an electrovibration stimulus at twice the input signal's frequency, i.e., 200~Hz. We selected this frequency as it is in the range of frequencies at which electrovibration is best perceived~\parencite{vardar2016perception}. The stimulus amplitude varied throughout the experiment, see the ``Procedure'' Section. The duration of the electrovibration stimulus was either 0.2 or 0.5 seconds, which was perceived twice in a trial that lasted 4 seconds by stroking the fingertip on the touchscreen back and forth in a left-right motion. The 0.2-second duration represented a 10~mm ridge (e.g., an edge or a small button) on the touchscreen, while the longer stimulus simulated a 25~mm patch (e.g., a small slider). Both stimuli were displayed such that their spatial center was aligned to the center of the touchscreen; see Figure S1 in Supplementary Materials for a schematic view of a trial with a 0.5-second stimulus. 

\subsubsection{Masking stimulus (whole-body motion)}
The masking stimulus consisted of mechanical vibrations generated by the SIMONA Research Simulator's motion system, affecting the whole body of the participant while interacting with the touchscreen inside the cabin. We selected vibrations simulating vertical turbulence accelerations on a flying vehicle as the masking stimulus. We tested two different types of turbulence signals, here referred to as Multisine and Gaussian, which were previously designed and tested by~\textcite{Khoshnewiszadeh2021680, Leto_2023} and had frequency components up to 10~Hz; see Supplementary Materials for details regarding their design.

The original Gaussian and Multisine turbulence signals as used in \parencite{Leto_2023} had a duration of 90~seconds. For the current experiment, 10 different 6-second segments of both signals were extracted for our stimulus interval (1-second fade-in, 4-second exposure, 1-second fade-out, see Figure S1. The 10 different masking stimuli were used to ensure a generalizable result was obtained for the electrovibration perception threshold.

\subsection{Procedure}
\label{section:procedure}
Before the experiments, each participant washed their hands with water and soap and dried them at the natural room temperature. Then, they read and signed the informed consent form. Afterward, each participant watched a safety video for the SIMONA Research Simulator, and they were briefed about experimental procedures and warned about motion sickness. Then, each participant sat in the simulator and arranged the seat to interact with the touchscreen comfortably. The touchscreen was also cleaned with alcohol before each session. Each experimental session was started when the participant pressed the trigger button on the side stick. 

The experiments aimed to determine the absolute detection thresholds of participants for the target (electrovibration) stimuli in the presence and absence of a masking stimulus (whole-body vibrations). The experiments followed the two-alternative-forced-choice (2AFC) method~\parencite{fechner1860AFC}. The stimuli were displayed in two temporal intervals, see Figure S1, which were signaled to the participant as red and green using a graphical user interface (GUI) designed in Python. Each interval lasted for 4~seconds. Only one of those intervals contained the target stimulus, while the masking stimulus was present in both intervals. The location of the target stimulus was randomized for each trial and participant. 

Each participant was instructed to hold the index finger of their dominant hand on the screen indicated with a virtual cursor and move in the tangential direction while synchronizing their finger movements with the cursor's motion, which moved with a speed of 50~mm/s. This finger speed resulted in an exploration area of 100~mm wide. In one interval, they moved one stroke of two seconds to the right and one stroke of two seconds to the left. After the red interval ended, the green interval started after a two-second gap, and the participant repeated the same procedure. 

After experiencing the two intervals, the participant's task was to indicate whether the target stimulus was in the red or green interval, for which they had unlimited time. They registered their choice by pressing the red or green buttons on the sidestick. When they were ready for the subsequent trial, they pressed the sidestick's trigger button. 

The amplitude of the target stimulus was modified via the three-up/one-down adaptive staircase method~\parencite{levitt1971adaptive}. Each session started with a stimulus generated by a (peak) input voltage of 50~V. This amplitude was chosen based on preliminary experiments and participant training sessions to ensure that all participants clearly felt the initial stimulus. If the participant gave three correct answers (not necessarily consecutive), the voltage amplitude was decreased by 5~dB. If the participant gave one incorrect answer, the amplitude was increased by 5~dB. The change from increasing intensity to decreasing and vice versa is called reversal. After one reversal, the step size was changed to 1~dB. The session ended after five reversals in a $\pm$1 dB level, and the mean value of these last five reversal voltage levels was taken as the absolute threshold. The dB unit used in research on electrovibration is described as $20\log_{10}(V_{p})$ with $V_{p}$ being the peak voltage of the touchscreen. The maximum peak amplitude of the applied voltage was set at 50~V. For all participants, the whole staircase was completed in approximately 30--50 trials.  

Each participant completed the experiments in six sessions: 2 test signal durations (0.2 and 0.5 s) $\times$ 3 whole-body vibration cases (no motion, Multisine, and Gaussian motion). The sessions were conducted in six different orders between participants to balance out order effects. During one trial of the experiment, the exact same turbulence stimulus was used for both intervals, see Figure S1, to enable an unbiased comparison between the two intervals. For different trials, different turbulence signals were used. 

The turbulence signal for each trial was a randomized choice from the ten 6-second stimuli that were available.

From the collected data, the measured absolute threshold voltages, average finger speeds, average applied forces, and average rate of change in applied forces were extracted; see Supplementary Materials for signal processing details. For the statistical analysis, we first checked the normality of the extracted data using Shapiro-Wilk tests. As one or more samples were not normally distributed for all data types, we used non-parametric Friedman ANOVAs to test the main effects. Then, we applied Bonferroni-corrected Wilcoxon signed rank tests for post-hoc pairwise comparisons. All data are depicted in box plots in this paper, with the results for the 0.2-second and 0.5-second electrovibration stimuli shown in green and yellow, respectively. The center red lines signify medians, while the box limits delineate the 25th and 75th percentiles. Whiskers extend to 1.5 times the interquartile range. Outliers are denoted by red plus signs (+), while yellow diamonds ($\diamond$) signify sample means. Circles represent ($\circ$) individual measurements from each participant. Overbraces with indicated asterisks indicate statistically significant differences (* for p$<$0.05 and ** for p$<$0.001).

\section{Results}

\begin{figure}[htb!]
    \centering
    \includegraphics[width=0.75\linewidth]{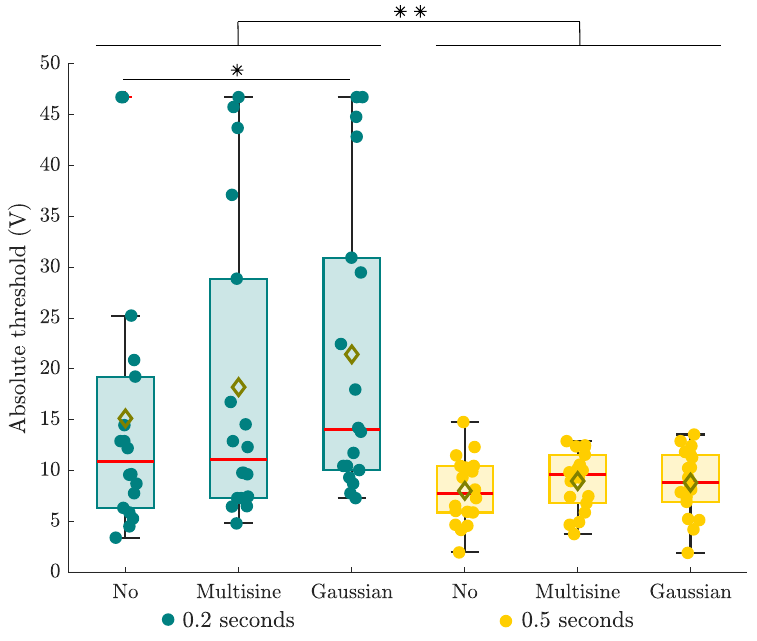}
    \caption{Absolute threshold voltages for all tested experiment conditions.
    }
    \label{fig:thresholdboxplot}
\end{figure}

The measured absolute threshold voltages of the participants are shown in Figure~\ref{fig:thresholdboxplot}. As seen here, the distribution for 0.2-second electrovibration stimulus without turbulence has two outliers, meaning that two participants (P6 and P8) already had very high thresholds (>45~V) without turbulence. With both turbulence cases, even more of these very high thresholds were measured, indicating that these participants could not reliably feel the stimulus. 
Overall, a statistically significant difference in the absolute threshold across all test cases was found $\chi^2(5) = 35.8$, p$<$0.001, as well as a highly significant effect of electrovibration duration ($\chi^2(1) = 25.8$, p$<$0.001). On average, across all turbulence conditions, the median thresholds were 38\% higher for the 0.2-second stimulus, see Figure~\ref{fig:thresholdboxplot}. While no significant effect of whole-body vibrations was found for the 0.5-second stimulus ($\chi^2(2) = 4.25$, p$>$0.05), the increased thresholds found with the Multisine and Gaussian turbulence were significant for the 0.2-second stimulus ($\chi^2(2) = 7.74$, p$<$0.05).  Pairwise comparisons also showed a significant difference between the no-vibration and Gaussian vibration conditions (p$<$0.05).

\begin{figure}[h!]
\captionsetup[subfigure]{justification=centering}
    \centering
    \begin{subfigure}{0.49\linewidth}
        \includegraphics[width=1\linewidth]{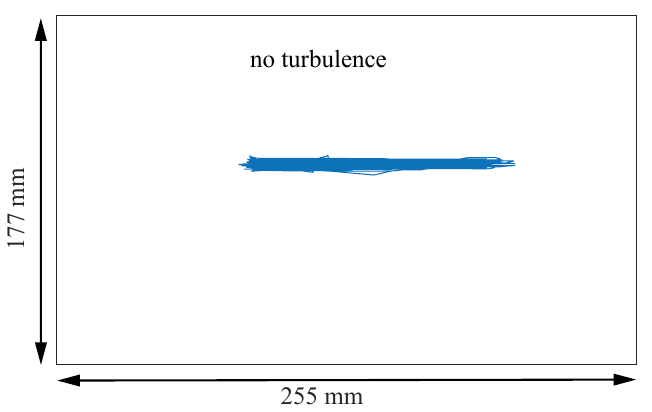}
        \caption{}
        \label{fig:fingerposcompa}
    \end{subfigure}
    \begin{subfigure}{0.49\linewidth}
        \includegraphics[width=1\linewidth]{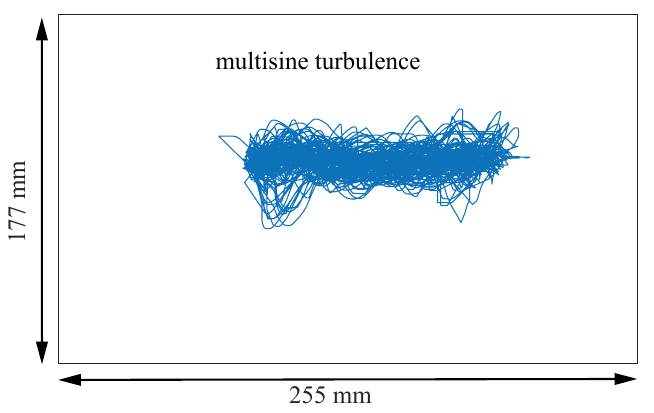}
        \caption{}
        \label{fig:fingerposcompb}
    \end{subfigure}
    \caption{Measured finger positions of a participant during two complete experiments with 0.5-second electrovibration stimuli during (a) no turbulence (47 trials) and (b) Multisine turbulence (37 trials).}
    \label{fig:fingerposcomp}
\end{figure}

Figure~\ref{fig:fingerposcomp} shows an example finger position measurement (34 aggregated trials) from one participant in the absence and presence of whole-body vibrations (compare Figure~\ref{fig:fingerposcomp}a and Figure~\ref{fig:fingerposcomp}b). The figure axes match the geometry and aspect ratio of the touchscreen, showing the participant trying to follow a line in the center of the screen where the cursor was moving. The effect of biodynamic feedthrough of the whole-body vibrations is visible, with an almost straight horizontal finger trajectory during no turbulence and large undesired vertical movements during the Multisine vibration.  

\begin{figure}[h!]
    \centering
    \includegraphics[width=0.75\linewidth]{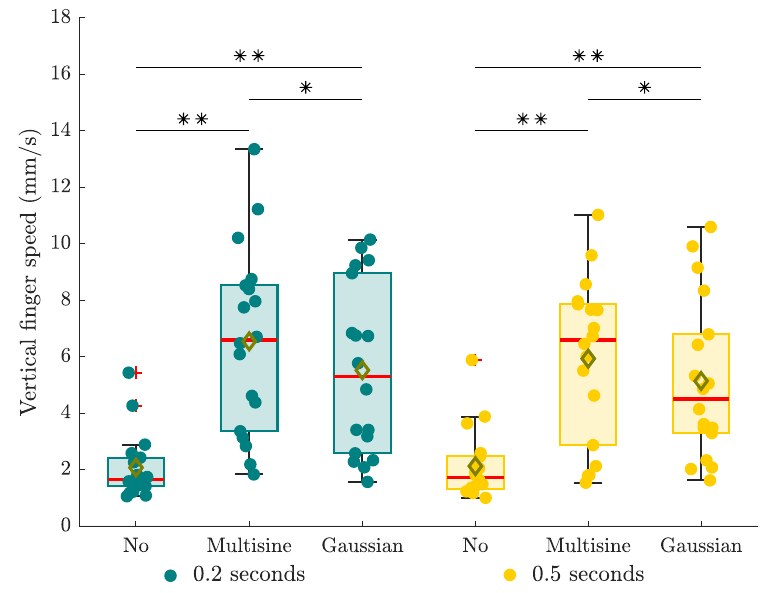}
    \caption{Average finger speeds of participants in the vertical direction for all tested experiment conditions.
    }
    \label{fig:boxplotymov}
\end{figure}

The average finger speeds in the vertical direction measured for the different experiment conditions are shown in Figure~\ref{fig:boxplotymov}. 
Overall, a highly significant variation in vertical finger speed was found across all tested conditions ($\chi^2(5) = 57.5$, p$<$0.001). We found that turbulence significantly increased the vertical finger speed for both stimulus durations (0.2 s: $\chi^2(2) = 28.8$, p$<$0.001, 0.5 s: $\chi^2(2) = 21.8$, p$<$0.001), whereas the target stimulus duration did not have an effect($\chi^2(1) = 3.63$, p$>$0.05).

Pairwise comparisons showed that there was a significant difference between the no vibration and both vibration conditions for both stimulus durations (p$<$0.01), but also between the two different vibration conditions for the 0.2-second stimulus (p$<$0.05). Figure~\ref{fig:boxplotymov} shows that the average vertical finger speed was indeed consistently around 25\% lower with the Gaussian vibration, which is explained by its reduced high-frequency power compared to the Multisine turbulence (see Supplementary Materials). We also analyzed the average horizontal finger speeds, but found no significant effect of either vibration type or electrovibration stimulus duration.

\begin{figure}[h!]
    \centering
    \includegraphics[width=0.75\linewidth]{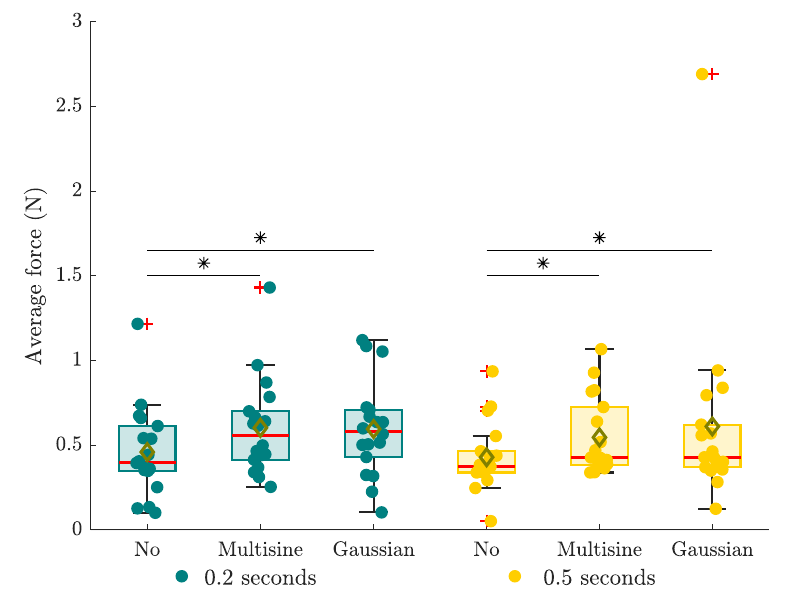}
    \caption{Average applied forces of participants for the tested experiment conditions.
    }
    \label{fig:boxplotforce}
\end{figure}

Figure~\ref{fig:boxplotforce} shows the average normal force applied on the touchscreen during the experiments.
The overall Friedman ANOVA showed a significant effect across all tested conditions ($\chi^2(5) = 13.8$, p$<$0.05). Furthermore, we found a significant effect of vibration on the average force for both the 0.2-second ($\chi^2(2) = 8.44$, p$<$0.05) and 0.5-second ($\chi^2(2) = 14.3$, p$<$0.001) target stimuli, but no main effect of stimulus duration ($\chi^2(1) = 0.67$, p$>$0.05. 
When considering pairwise comparisons, only the no-vibration condition was found to be significantly different from both vibration conditions for both stimulus durations (p$<$0.05), due to average normal forces increases of 0.17 N and 0.05 N for the 0.2 and 0.5-second stimulus durations, respectively. Overall, this indicates that the presence of Multisine or Gaussian turbulence resulted in elevated forces being applied to the screen, an effect that is stronger for the shorter 0.2-second stimulus, see Figure~\ref{fig:boxplotforce}.

\begin{figure}[t!]
    \centering
    \includegraphics[width=0.75\linewidth]{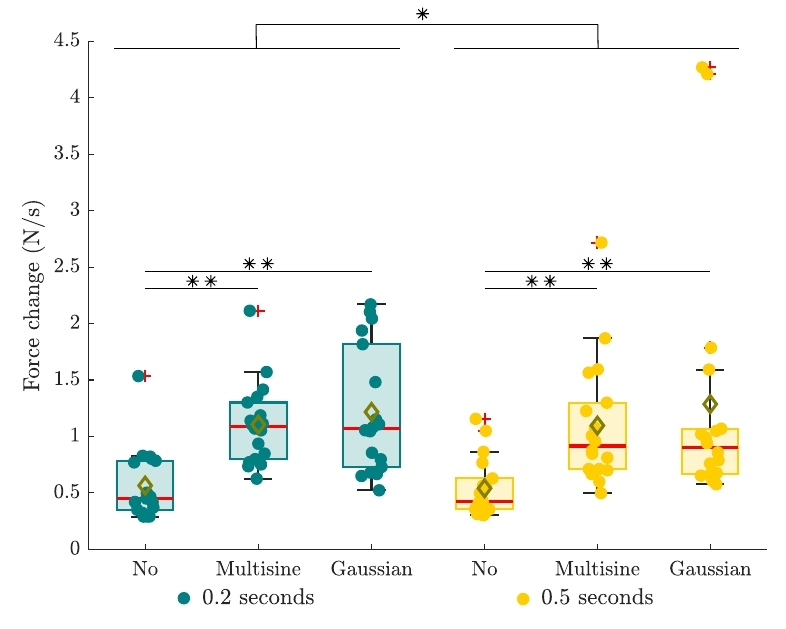}
    \caption{Average rate of change in the average applied forces of participants across different experimental conditions.
    }
    \label{fig:boxplotforcediff}
\end{figure}

The calculated rate of change in applied force for different experimental conditions is shown in Figure~\ref{fig:boxplotforcediff}.
The average force change varied significantly across all experiment conditions ($\chi^2(5) = 63.6$, p$<$0.001), due to a more than 100\% increase in the presence of whole-body vibrations for both the 0.2-second stimulus ($\chi^2(2) = 24.3$, p$<$0.001) and the 0.5-second stimulus $\chi^2(2) = 27.1$, p$<$0.001. Pairwise comparisons showed that this increase in average force change was not significantly different between the Multisine and Gaussian turbulence cases for both stimulus durations (p$>$0.05).
A weaker effect of stimulus duration was also found ($\chi^2(1) = 6.00$, p$<$0.05), due to 0.17 N/s lower median values for the 0.5-second stimulus with whole-body vibrations. Pairwise comparisons, however, showed no statistically significant differences for the individual turbulence conditions.

\section{Discussion and Conclusions}
This study investigates the influence of whole-body vibrations, such as aircraft turbulence, on the perception of electrovibration displayed on touchscreens. For this objective, we measured the absolute detection thresholds of 18 human participants for electrovibration stimuli with varying durations, both in the absence and presence of two types of whole-body vibrations. Concurrently, we measured participants' applied normal force and finger speed during the experiments. We hypothesized that the detection thresholds for shorter-duration electrovibration stimuli would surpass those for longer durations. Furthermore, we anticipated an increase in detection thresholds, finger speed in the vertical direction, applied normal force, and the rate of change of this force when whole-body vibrations were introduced.  

The first hypothesis, predicting an increase in absolute thresholds for shorter electrovibration stimulus duration, is supported by the significantly higher thresholds observed with shorter electrovibration durations compared to longer ones (see Figure~\ref{fig:thresholdboxplot}, green and yellow box plots). This outcome can be attributed to a temporal summation, i.e., the integration of energy over time by a sensory system; here the Pacinians, the corresponding mechanoreceptors for perceiving 200~Hz electrovibration stimuli at the threshold level~\parencite{vardar2017waveform}. Previous studies conducted with 200~Hz vibrotactile stimuli also demonstrated a similar behavior, where the perceptual thresholds decreased as a function of stimulus duration~\parencite{geschieder1982temporal, checkosky1994duration}.  

Contrary to the second hypothesis suggesting an increase in absolute thresholds due to turbulence, we only found significantly elevated measured thresholds with turbulence for the 0.2-second electrovibration stimulus (see Figure~\ref{fig:thresholdboxplot}). This result indicates the absence of perceptual masking, possibly explained by the considerable frequency difference between the turbulence and electrovibration stimuli. Past studies on vibrotactile stimuli revealed that perceptual masking is most pronounced when test and masking stimuli share a similar or close frequency~\parencite{hamer1983masking}. Similar findings were observed when stimuli were applied at the same or distant sites~\parencite{verrillo1982site}. Previous research~\parencite{ryu_mechanical_vibration, jamalzadeh2019remote} corroborated the influence of mechanical vibration on electrovibration perception, particularly when both stimuli shared similar frequencies~\parencite{ryu_mechanical_vibration}. These results confirm that the masking effect diminishes with higher frequency differences.

The significantly elevated thresholds with turbulence for shorter electrovibration stimuli can be attributed to significant force changes during turbulence (see next paragraph) combined with more erratic finger movements, see Figure~\ref{fig:boxplotforcediff} and Figure~\ref{fig:boxplotymov}, respectively. Participants likely often missed the short 0.2-second stimulus during turbulence, e.g., as in case of simultaneous normal force or finger speed variations. In contrast, a more extended stimulus duration gave participants greater chances to perceive the electrovibration despite sudden touch interaction changes. The increased number of exceedingly high thresholds ($\geq$40~V) for the 0.2-second stimulus during turbulence provides evidence for this interpretation. Considering that the maximum allowed voltage amplitude was 50~V, these participants simply felt no consistent tactile stimuli during turbulence. Furthermore, several participants commented about their fingers ``jumping'' over the screen during turbulence, supporting this notion. Unfortunately, these `jumps' could not be reliably extracted from the IR or force sensor signals due to high noise levels. Future research could employ electrical impedance measurements to detect such instances of loss-of-contact~\parencite{forte2024selftouch, vardar2021motion}. 

Our hypothesis asserting that the vertical finger scan speed (i.e., the finger speed in the direction of turbulence motion) would significantly increase during turbulence is confirmed (see Figure~\ref{fig:boxplotymov}). The increase from average finger speeds around 2~mm/s to around 6~mm/s, caused by additional involuntary finger movements as a result of the applied vertical whole-body vibration, is consistent with earlier experiments on touchscreen biodynamic feedthrough \parencite{mobertz2018bdft, Khoshnewiszadeh2021680}. Additionally, we observed that the average vertical finger speed was 25\% lower with the Gaussian turbulence than the Multisine turbulence. This difference can be explained by the frequency-dependent power distribution of the turbulence signals; the Multisine signal caused comparatively stronger vibrations for the higher frequencies in the masking stimulus, causing higher vertical finger speeds. 

Lastly, the hypothesis dictating that turbulence would significantly increase the applied normal force and its rate of change is also supported by our data, see Figures~\ref{fig:boxplotforce} and \ref{fig:boxplotforcediff}. The elevated applied normal force during turbulence likely results from deliberate compensation for biodynamic feedthrough, leading participants to exert more (steady-state) force on the touchscreen to prevent involuntary sliding motions induced by the turbulence~\parencite{venrooij2013bdft, Khoshnewiszadeh2021680}. The doubling of the derivative of the normal force in the presence of turbulence can be explained as a direct result of acceleration feedthrough. Equivalent to how involuntary movements in the participant's arm and hand cause an additional perturbation component in the on-screen finger movement~\parencite{mobertz2018bdft, Khoshnewiszadeh2021680}, they also add a similar component to the rate of change of the normal force.

Interestingly, despite the significant changes in the finger contact dynamics (e.g., changes in contact force and finger speed) during turbulence, there were no substantial alterations in the perceptual thresholds for the long-duration electrovibration stimulus. Nonetheless, past research has shown that changes in finger speed and applied normal force affect finger contact area, the air gap between the touchscreen, and the stick-slip behavior of the fingertip, directly influence the generated electrostatic force and its perception~\parencite{vardar2021motion, vardar2017waveform}. Some of these conditions are known to adversely affect each other, such as increased applied force leading to increased contact area~\parencite{ayyildiz2018contact}, but increased stick-slip behavior and a decreased air gap; consequently, the overall effect could hide such underlying changes. A more detailed investigation of how these factors dynamically change during turbulence is needed, also taking into account that these physical factors regarding finger contact dynamics are known to vary greatly between people~\parencite{serhat2022contact, nam2020stickiness}. Finally, in our experiment the electrovibration stimulus was presented in a time-based manner; hence it could be perceived anywhere on the screen, regardless of the actual finger movement. For more realistic presentation of the electrovibration only at a fixed location on the graphical interface, the effects of turbulence motion are likely more severe. This is the next crucial hypothesis that needs to be tested in future work. 

Our results offer valuable insights for the future design of touchscreen interactions in vehicle cockpits. Electrovibration-rendered virtual buttons or sliders should always be substantial enough to generate stimuli that are larger than 1 cm (0.2 s) in size to minimize susceptibility to turbulence. Furthermore, the finger speed and normal force fluctuations induced by turbulence likely need to be actively countered in tactile rendering to avoid undesired variations in electrovibration intensity.

\section{Key Points}
\begin{itemize}
\item{Absolute electrovibration detection thresholds are 38\% higher for short (0.2-second) electrovibration stimuli compared to longer stimulus durations (0.5 seconds).}
\item{Low-frequency whole-body vibrations such as turbulence impair the perception of short (0.2-second) electrovibration stimuli, but not longer ones (0.5 seconds).}
\item{Perceptual masking does not occur for electrovibration stimuli in the presence of low-frequency whole-body vibrations.}
\item{Involuntary arm movements due to biodynamic feedthrough of whole-body vibrations increase the finger scan speed, the average applied force, and the rate of force change.}
\end{itemize}

\printbibliography

\section{Biographies}

Jan D.~A. Vuik completed his M.Sc. degree in Robotics at Delft University of Technology (The Netherlands) in 2023. He is currently a Junior Data Scientist at Lely Technologies. \\

Daan M. Pool is an assistant professor at the Department of Control and Operations of the Faculty of Aerospace Engineering at Delft University of Technology. He obtained his Ph.D. degree in Aerospace Engineering at the same institution in 2012. \\

Yasemin Vardar is an assistant professor at the Department of Cognitive Robotics of the Faculty of Mechanical Engineering at Delft University of Technology. She earned her Ph.D. degree in Mechanical Engineering at Ko\c{c} University (Turkey) in 2018.

\section{Supplementary materials}
\begin{figure}[htb!]
    \centering
        \renewcommand{\figurename}{Figure S1}
        \renewcommand{\thefigure}{}
    \includegraphics[width=0.75\linewidth]{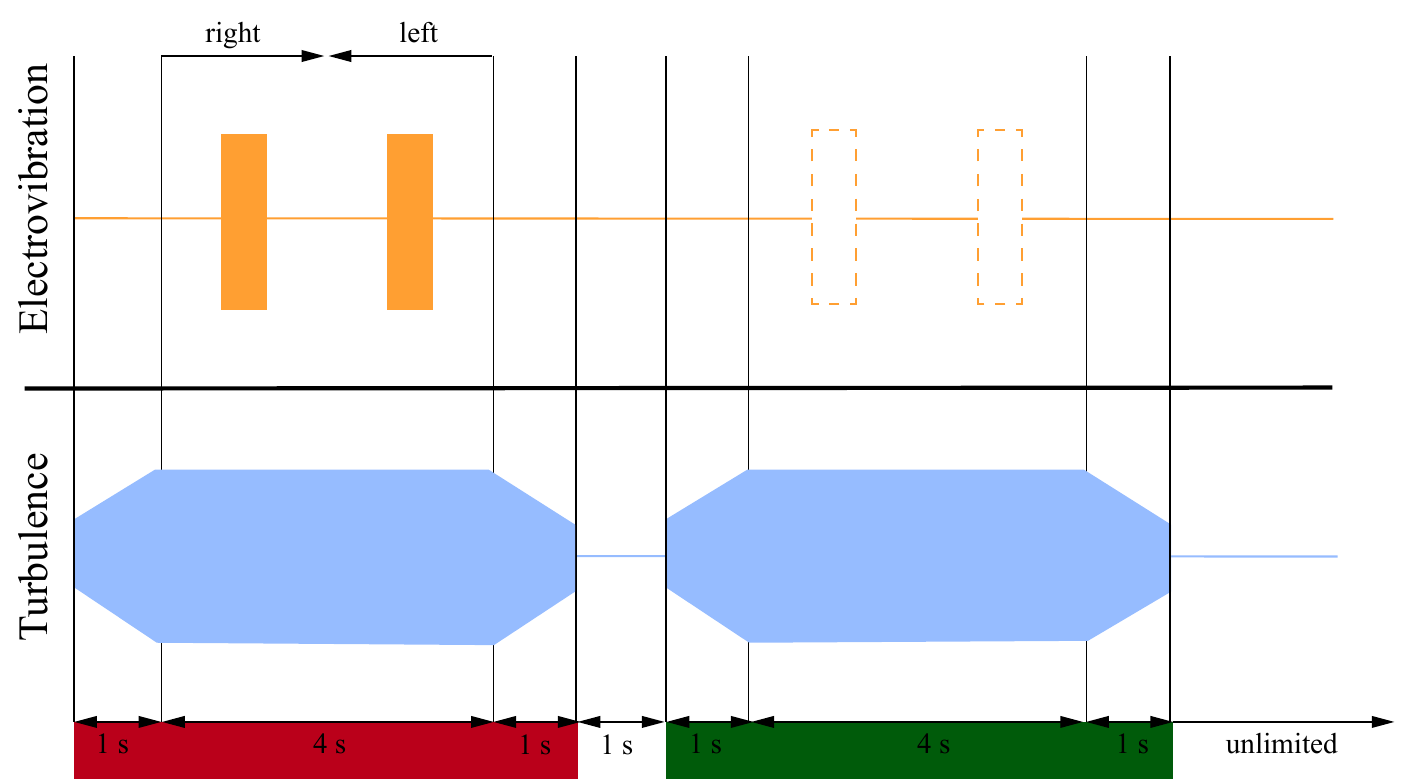}
    \newline
    \caption{Example stimulus timing diagram for the absolute threshold experiments. The target (electrovibration) stimulus was generated by bursts of (100~Hz) input voltage signals applied to the touchscreen. In threshold experiments with masked stimuli, the masking (turbulence) stimulus was whole-body vibrations with one-second fade-in and fade-out. Both target and masking stimuli were displayed in two temporal intervals, which were signaled to the participants as red and green. The electrovibration stimulus was displayed randomly at either the red or green intervals. In the interval which did not have the electrovibration stimulus, the participants explored the smooth glass surface. In each interval, the participants explored the touchscreen in two strokes with a scan speed of 50~mm/s; each lasted 2~seconds. The participants gave their responses after the green interval ended.}
    \label{fig:trialscheme}
\end{figure}

\subsection{Design of masking stimuli}
Both turbulence signals were designed for an RMS vertical acceleration of 0.75~m/s\textsuperscript{2}, representative of heavy turbulence \parencite{Leto_2023, Coutts2019}.

The Multisine turbulence signal was identical to those used in previous experiments \parencite{mobertz2018bdft, Khoshnewiszadeh2021680} and defined as a sum of ten sinusoids: 

\begin{equation}
    \label{eq:multisine}
    \sum_{k=1}^{10}A_{k}\sin(\omega_{k}t + \phi_{k}),
\end{equation}

\noindent where $A_{k}$, $\omega_{k}$, and $\phi_{k}$ represent the amplitude, frequency, and phase offset of each sinusoid. The frequencies $\omega_{k}$ ranged from 0.06~Hz (0.383~rad/s) to 2.76~Hz (17.33~rad/s). For the Multisine vibrations, 50\% and 100\% of the vertical acceleration RMS is due to components below 1.67~Hz and 2.76~Hz, respectively; i.e., much lower frequencies than the 100~Hz of our target stimulus.

The Gaussian turbulence represented the vertical accelerations of a simulated Cessna Citation 500 business jet \parencite{VanDerLinden1996} flying through stationary turbulence, modeled using Dryden spectra \parencite{VandeMoesdijk1978, Leto_2023}. For the Gaussian turbulence, 50\% of the vertical acceleration RMS is due to components below 0.52~Hz, i.e., more low-frequency power compared to the Multisine turbulence. Furthermore, 96.4\% of the signal's power was at frequencies below 2.76~Hz (the maximum sinusoid frequency for the Multisine turbulence). To protect the motion platform's hardware from undesired low-amplitude high-frequency vibrations, the Gaussian turbulence signal was filtered with a second-order low-pass filter with a cut-off frequency of 10~Hz.

\subsection{Data analysis}
We stored the data acquired from the experiment in one file per staircase and analyzed it with a Matlab program. The force and finger position data consisted of a time series per staircase, with parts of the breaks present. We extracted the trial data from these time series by taking only the data from the start and finish of every interval. Any data from either the breaks between trials or the breaks between intervals was omitted. The force data was recorded at 2000 Hz, the maximum achievable data rate of the sensors. The finger position data was recorded at 100 Hz, the locked frequency the IR sensor records. After calibrating the force sensors, we calculated the average applied normal force using the force from every sensor and the finger location to make a weighted average. A zero-phase digital second-order low-pass Butterworth filter filtered the force data to reduce the effect of high-frequency measurement noise, especially for the force change signal (time derivative of the force measurement). The finger location data was converted from pixel to millimeters, after which the data points outside the screen were filtered out. Then, the x- and y-location time series were differentiated to obtain the finger speed in both directions. In this data, values higher than 1000~mm/s, which corresponded to instances of loss-of-contact with the screen, were filtered out.

\end{document}